\documentstyle[sprocl]{article}

\input{psfig}
\def\Journal#1#2#3#4{{#1} {\bf #2}, #3 (#4)}

\def\PRL{\em Phys. Rev. Lett.}
\def\PRD{{\em Phys. Rev.} D}

\def\ra{\rightarrow}

\def\be{\begin{equation}}
\def\ee{\end{equation}}
\def\bea{\begin{eqnarray}}
\def\eea{\end{eqnarray}}

\newcommand{\SC}{\langle{\cal O}_8^{\psi '}(^1S_0)\rangle}

\newcommand{\PS}{\langle{\cal O}_8^{\psi '}(^3P_0)\rangle} 

\begin{document}

\title{PHOTOPRODUCTION OF {\large\boldmath $\psi^{\prime}$} 
AS A TEST OF \\COLOR--OCTET MODEL}

\author{K. SUDOH}
\address{Graduate School of Science and Technology, Kobe University\\
1--1, Rokkodai, Nada, kobe, 657--8501, Japan\\
E--mail: sudou@radix.h.kobe--u.ac.jp} 

\author{T. MORII and D. ROY}
\address{Graduate School of Science and Technology, Kobe University}

\maketitle\abstracts{\hspace*{0.25in}
We have calculated the cross section of $\psi^{\prime}$ photoproduction 
at small--$p_{T}$ regions in polarized $\gamma p$ collisions for testing 
the color--octet model, which is based on NRQCD factorization formalism. 
We show that the measurement of the two--spin asymmetry of this process 
might be effective to constrain the value of NRQCD matrix elements.} 

\section{Introduction}
\hspace*{0.25in}
Recently, it has been reported that the cross sections of prompt $J/\psi$ 
and $\psi^{\prime}$ production in unpolarized $p\bar{p}$ collisions 
measured by the CDF collaboration are largely inconsistent with the 
conventional color--singlet model calculation \cite{Abe92}. 
Far from that, these dramatic discrepancies could be explained well by 
taking a new color--octet model \cite{Cho95} into account. 
A rigorous formulation of the color--octet model has been presented in 
terms of a beautiful effective field theory called nonrelativistic QCD 
(NRQCD), in which the ${\cal{O}}(v)$ corrections of a relative velocity 
between the bound heavy quarks can be systematically calculated 
\cite{Bodwin95}. 
Physics of the color--octet model is now one of the most interesting topics 
for heavy quarkonium production at high energy. 

In this work, as another test of the color--octet model we propose a 
different process, 
\be
\vec{\gamma} + \vec{p} \ra \psi^{\prime} + X , 
\label{eq:process}
\ee
at small--$p_{T}$ regions which might be observed in the forthcoming HERA 
experiment, where both of an incident $\gamma$ and target $p$ are 
longitudinally polarized. 
The process is dominated by the $s$--channel photon--gluon fusion, so that 
the color--octet contribution is dominated at small--$p_{T}$ regions. 
Hence we consentrate on only two lowest diagrams here. 
Note that in the case of $J/\psi$ production, there are some additional 
processes like $\psi^{\prime} \ra J/\psi + X$, in addition to an original 
direct production of $J/\psi$, therefore the analysis must be complicated. 
Furthermore, the cross section and two--spin asymmetry are very sensitive 
to the gluon density in a proton, since the $\psi^{\prime}$ is 
dominantly produced by photon--gluon fusion. 
Thus one can get good information on the spin--dependent gluon distribution 
in a proton by analyzing the polarized process like this \cite{Sudoh98}. 

\section{Two--Spin Asymmetry for 
{\boldmath $\gamma p \ra \psi^{\prime}X$}}
\hspace*{0.25in}
Let us introduce a two--spin asymmetry $A_{LL}$ defined as 
\be
A_{LL} \equiv \frac{\left[d\sigma_{++}-d\sigma_{+-}+
d\sigma_{--}-d\sigma_{-+}\right]}
{\left[d\sigma_{++}+d\sigma_{+-}+
d\sigma_{--}+d\sigma_{-+}\right]} = \frac{d\Delta\sigma}{d\sigma}~,
\label{eq:A_LL} 
\ee
where $d\sigma_{+-}$, for instance, denotes that the helicity of one beam 
particle is positive and the other is negative. 

The spin--independent and spin--dependent total cross sections via the 
color--octet state are given by \cite{Gupta97} 
\be
\sigma(\gamma p \ra \psi^{\prime}) = \left.\frac{\pi^{3}e_{c}^{2}\alpha_{em}
\alpha{s}}{4m_{c}^{5}} xg(x,Q^{2}) \right |_{x = 4m_{c}^{2}/s}
\left\{ \SC + \frac{7}{m_{c}^{2}}\PS \right\} ,
\label{eq:SIC}
\ee
\be
\Delta\sigma(\gamma p \ra \psi^{\prime}) = \left.\frac{\pi^{3}e_{c}^{2}
\alpha_{em}\alpha{s}}{4m_{c}^{5}} x\Delta g(x,Q^{2}) \right |_{x = 
4m_{c}^{2}/s} \left\{ \SC - \frac{1}{m_{c}^{2}}\PS \right\} ,
\label{eq:SDC}
\ee
where $g(x,Q^{2})$ and $\Delta g(x,Q^{2})$ are spin--independent and 
spin--dependent gluon distribution functions with the momentum fraction $x$ 
at any $Q^{2}$, respectively. 
The factors $\SC$ and $\PS$ represent nonperturbative long distance parameters 
of NRQCD matrix elements and are associated with the production of a 
$c\bar{c}$ pair in a color--octet $^{1}S_{0}$ and $^{3}P_{0}$ states, 
respectively. 
The recent analysis on charmonium hadroproductions results in 
$\SC + (7/m_{c}^{2})\PS \approx 5.2\times 10^{-3} {\rm [GeV^{3}]}$ 
\cite{Beneke96} and 
$\frac{1}{3}\SC + (1/m_{c}^{2})\PS \approx (5.9\pm 1.9)\times 10^{-3} 
{\rm [GeV^{3}]}$ \cite{Leibovich97}, 
which lead to 
\be
\frac{\tilde{\Theta}}{\Theta} \equiv \frac{\SC-\frac{1}{m_{c}^{2}}\PS}
{\SC+\frac{7}{m_{c}^{2}}\PS} \approx 3.6 \sim 8.0 . 
\label{eq:ratio}
\ee
Using this value and various parton distributions, we have numerically 
calculated the two--spin asymmetry $A_{LL}$ originated from the color--octet 
state. 
Note that the $A_{LL}$ in Eq. (\ref{eq:A_LL}) has a simple form 
\be
A_{LL}^{\psi^{\prime}} (\gamma p) = \frac{\Delta g(x,Q^{2})}{g(x,Q^{2})} 
\cdot \frac{\tilde{\Theta}}{\Theta} , 
\label{eq:A_LL2}
\ee
and is directly propotional to the ratio of gluon densities and of NRQCD 
matrix elements in this process. 

\section{Numerical Results}
\hspace*{0.25in}
We have calculated the $A_{LL}$ at relevant HERA energies, which are 
presented in figures. 
Since the value of $A_{LL}$ is rather large, we can sufficiently test the 
color--octet model in this reaction. 

Let us discuss the sensitivity on our results. 
In order to examine the experimental accuracy of the forthcoming HERA 
experiments, we have estimated the experimental sensitivity of the $A_{LL}$ 
for 100--day experiments at various $\sqrt{s}$ in the manner of 
Nowak \cite{Nowak}, using the expected data of beam or target polarization 
($P_{B}, P_{T} \sim 70\%$), the integrated luminosity (${\cal{L}\cdot T} \sim 
66 {\rm pb^{-1}}$), and the combined trigger and reconstruction efficiency 
($C \sim 50\%$) together with the values of unpolarized cross sections. 
As a result we found that the experimental sensitivity $\delta A_{LL}$ 
is order of magnitude $\sim 10^{-3}$, which is very small. 
Hence we can neglect it safely and our predictions are expected to be 
actually tested in the HERA experiments. 

However we can see in figures that the $A_{LL}$ goes over 1 for small 
$\sqrt{s}$ or large $x$ regions and become unphysical. 
To confirm that our results is reasonable, we attempt to explain such a 
odd behavior of the $A_{LL}$ below. 
We have already seen that the $A_{LL}$ is given by a product of 
$\Delta g(x)/g(x)$ and $\tilde{\Theta}/\Theta$. 
In the case of GS96 or GRSV96 parametrization which are widely used, 
the maximum value of $\Delta g(x)/g(x)$ is $0.2 \sim 0.35$, whereas the 
value of $\tilde{\Theta}/\Theta$ is $3.6 \sim 8.0$. 
If we rely on the value of the ratio of gluon distributions, the maximum 
value of the ratio of NRQCD matrix elements should be 
\be
\frac{\tilde{\Theta}}{\Theta} [{\rm max}] \leq 5.0 , 
\label{eq:max}
\ee
from the requierment that the $A_{LL}$ should be less than 1. 
Actually, the uncertainty of matrix elements seems to be larger than 
the one of gluon distribution, because the value of matrix elements obtained 
from the Tevatron data does not include the contributions of higher order 
QCD corrections. 
Therefore we might be able to constrain the value of matrix elements by 
measuring the two--spin asymmetry in polarized $\gamma p$ collisions. 

\section{Summary and Discussion}
\hspace*{0.25in}
We have proposed the photoproduction of $\psi^{\prime}$ at small--$p_{T}$ 
regions in polarized $\gamma p$ collisions which might be available in the 
forthcoming HERA experiments. 
We have calculated two--spin asymmetry $A_{LL}$ for various parameter 
regions $\tilde{\Theta}/\Theta = 3.6 \sim 8.0$, and found that the $A_{LL}$ 
is rather large in the regions $\sqrt{s}=10\sim 50 {\rm GeV}$, 
$x = 0.005\sim 0.1$. 
Therefore we conclude that we can sufficiently test the color--octet 
model in this process. 
In addition, we can constrain the value of NRQCD matrix elements though 
there still remain big uncertainties for $\Delta g(x)$. 
\noindent
\begin{figure}[t]
\parbox[t]{0.46\textwidth}
{
\begin{center}
\psfig{figure=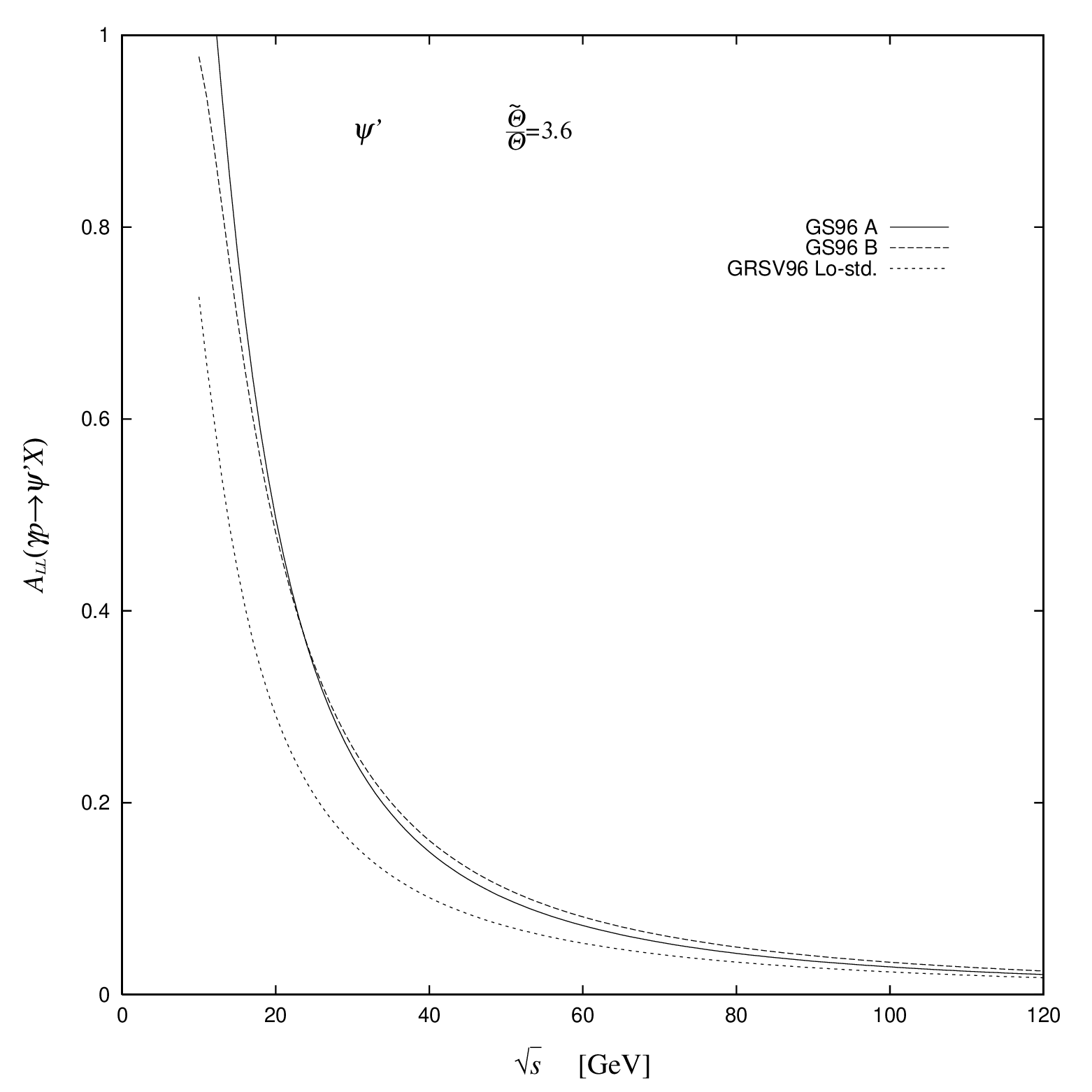,height=2in}
\end{center}
}
\hfill
\parbox[t]{0.46\textwidth}
{
\begin{center}
\psfig{figure=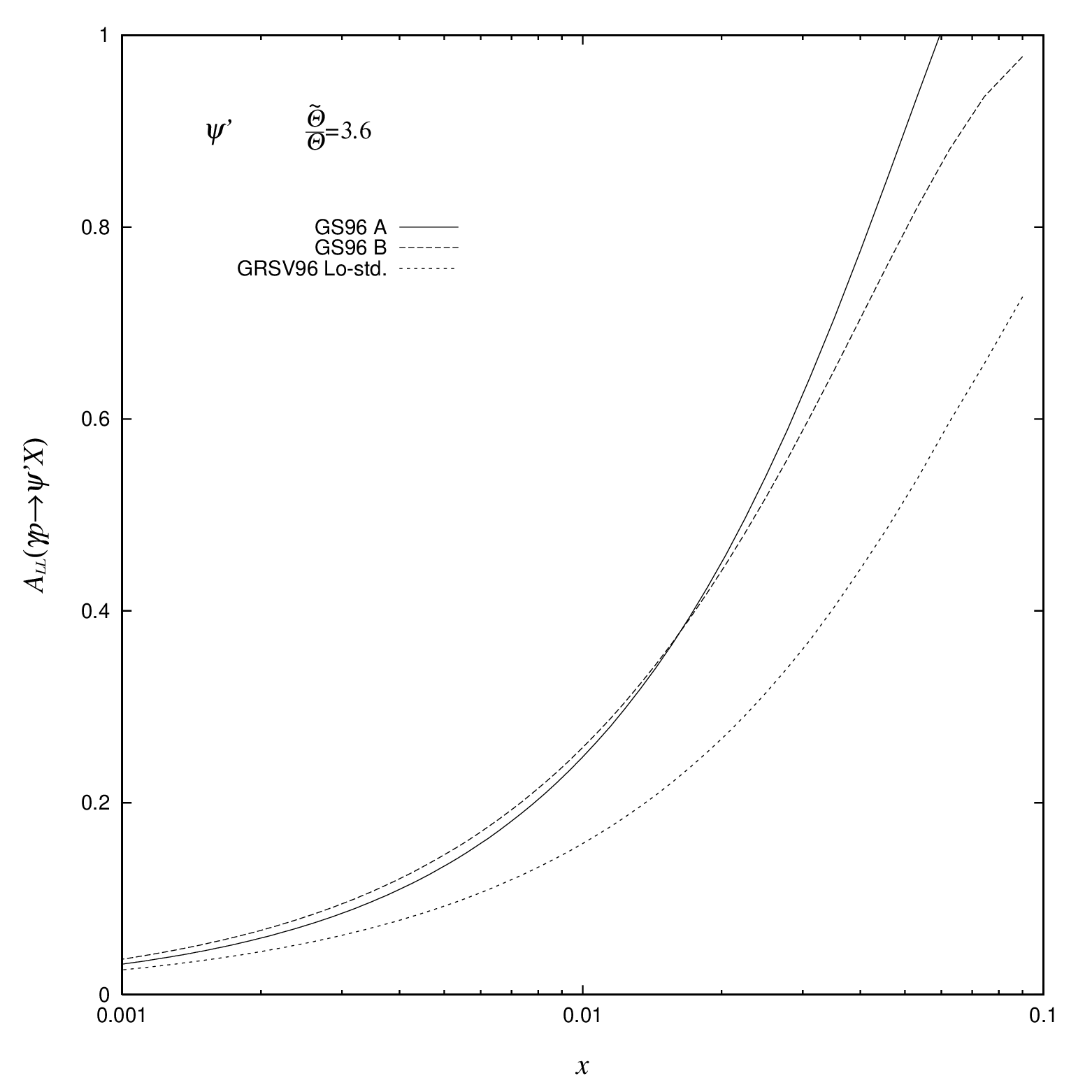,height=2in}
\end{center}
}
\vspace{-0.4cm}
\caption{The two--spin asymmetry $A_{LL}(\gamma p \ra \psi^{\prime}X)$ 
with the parameter $\tilde{\Theta}/\Theta = 3.6$ as a function of 
$\sqrt{s}$ and momentum fraction $x$. 
The solid, dashed and dotted lines show the case of set A of ref.[9], set B 
of ref.[9] and the 'standard scenario' of ref.[10], respectively. 
\label{fig:all}}
\end{figure}  

\end{document}